\documentclass[fleqn,twoside]{article}
\usepackage{espcrc2}

\usepackage{graphicx}

\title{Baryons and Baryonic Matter 
in Holographic QCD from Superstring}

\author{
Hideo Suganuma\address[kyoto]{
Department of Physics, Kyoto University, 
Kitashirakawa, Sakyo 606-8502, Japan},
Kanabu Nawa\address[osaka]{
Research Center for Nuclear Physics (RCNP), Osaka University, Ibaraki, Osaka 567-0047, Japan}
and
Toru Kojo\address[bnl]{
RIKEN BNL Research Center, 
Brookhaven National Laboratory, Upton, NY 11973, USA}
}

\begin{document}

\begin{abstract}
We study baryons and baryonic matter in holographic QCD 
using a D4/D8/$\overline{\rm D8}$ multi-D-brane system 
in the superstring theory. 
We obtain the chiral soliton solution for baryons 
in the four-dimensional meson theory 
derived from the multi-D-brane system. 
For the analysis of finite baryon-density matter, 
we investigate the chiral soliton on $S^3$ in holographic QCD, 
and find the delocalization of the soliton, 
i.e., the swelling of baryons in dense matter.
\end{abstract}

\maketitle

\section{The Yang-Mills theory from D-branes}

To understand the nonperturbative properties of 
QCD is one of the most important and difficult problems remaining 
in theoretical physics. 
As a recent progress, a new method called holographic QCD 
has been developed to analyze nonperturbative QCD 
using the superstring theory.

The superstring theory is well-defined in ten-dimensional space-time, 
and includes D$p$-branes \cite{P95} as ($p$+1)-dimensional soliton-like 
objects of fundamental strings. 
On the surface of $N$ D-branes, 
there exists U($N$) gauge symmetry and 
U($N$) gauge field appears from the open string.
In the space around the massive D-brane, a supergravity field is 
created.
Since the gravity field depends on the distance from the D-brane, 
one more coordinate appears in the gravity side.

In general, D-branes lead to 
SUSY theories, reflecting superstring nature. 
Following Witten \cite{W98}, to get non-SUSY gauge theories, 
we break SUSY explicitly by spatial $S^1$-compactification of D-branes 
with the periodic/anti-periodic boundary condition for 
bosons/fermions. The inverse radius of 
the $S^1$ is called as the Kaluza-Klein mass $M_{\rm KK}$, 
and the gaugino mass becomes $O(M_{\rm KK})$. 
Then, non-SUSY gauge theories are formed  
on the compactified D-brane at larger scale than 1/$M_{\rm KK}$. 

The four-dimensional non-SUSY U($N_c$) Yang-Mills theory is 
realized on $S^1$-compactified $N_c$ D4-branes, 
where only the gauge field ${\cal A}^\mu$ remains to be massless. 
On the other hand, the effect of the D4-brane can be also described by 
the gravity field around it, under the hypothesis of gauge/gravity 
correspondence \cite{M98}.
In fact, the Yang-Mills theory or the gauge sector of QCD is 
constructed using the $S^1$-compactified $N_c$ D4-branes, 
and it is transferred into 
a higher-dimensional gravity theory in the holographic framework.   
Due to the strong-weak coupling duality, 
nonperturbative quantities of large-$N_c$ QCD are 
calculable with the classical gravity theory \cite{M98}. 

\section{Construction of holographic QCD}

Four-dimensional massless QCD can be constructed 
with the D4/D8/$\overline{\rm D8}$ multi-D-brane system \cite{SS05} 
consisting of spatially $S^1$-compactified $N_c$ D4-branes 
attached with $N_f$ D8-$\overline{\rm D8}$ pairs, 
as shown in Fig.1(a). From this multi-D-brane system, the 
higher-dimensional theory of holographic QCD is formulated.

In holographic QCD, ``color'' and ``flavor'' are described 
as different physical objects, i.e., different D-branes: 
the D4 gives color and the D8 gives flavor. 
Here, gluons appear on the D4-brane, 
and quarks, which have color and flavor, 
appear at the cross point between D4 and D8/$\overline{\rm D8}$.

\begin{figure}[h]
\begin{center}
       \includegraphics[width=7cm]{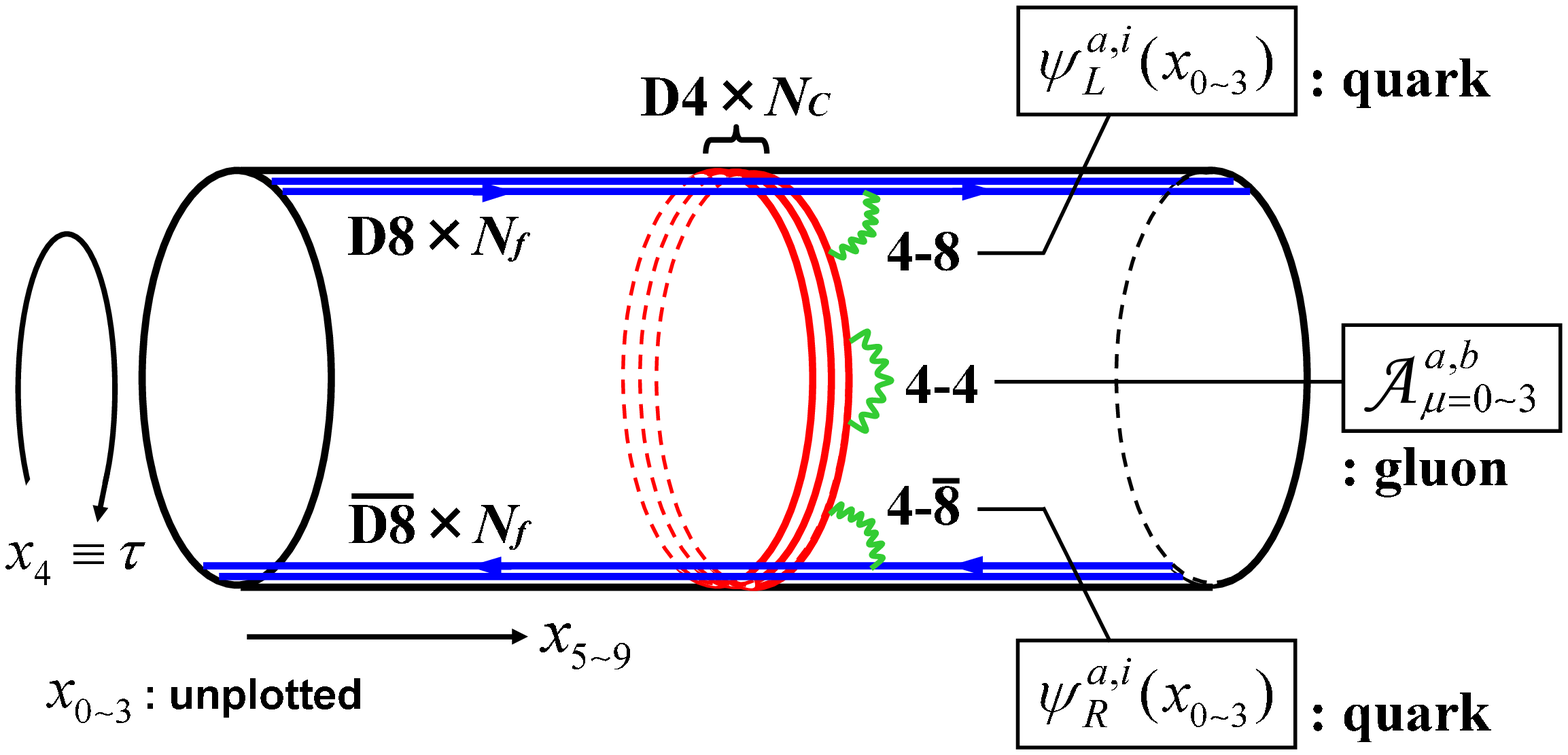}\\
\vspace{0.5cm}
       \includegraphics[width=7cm]{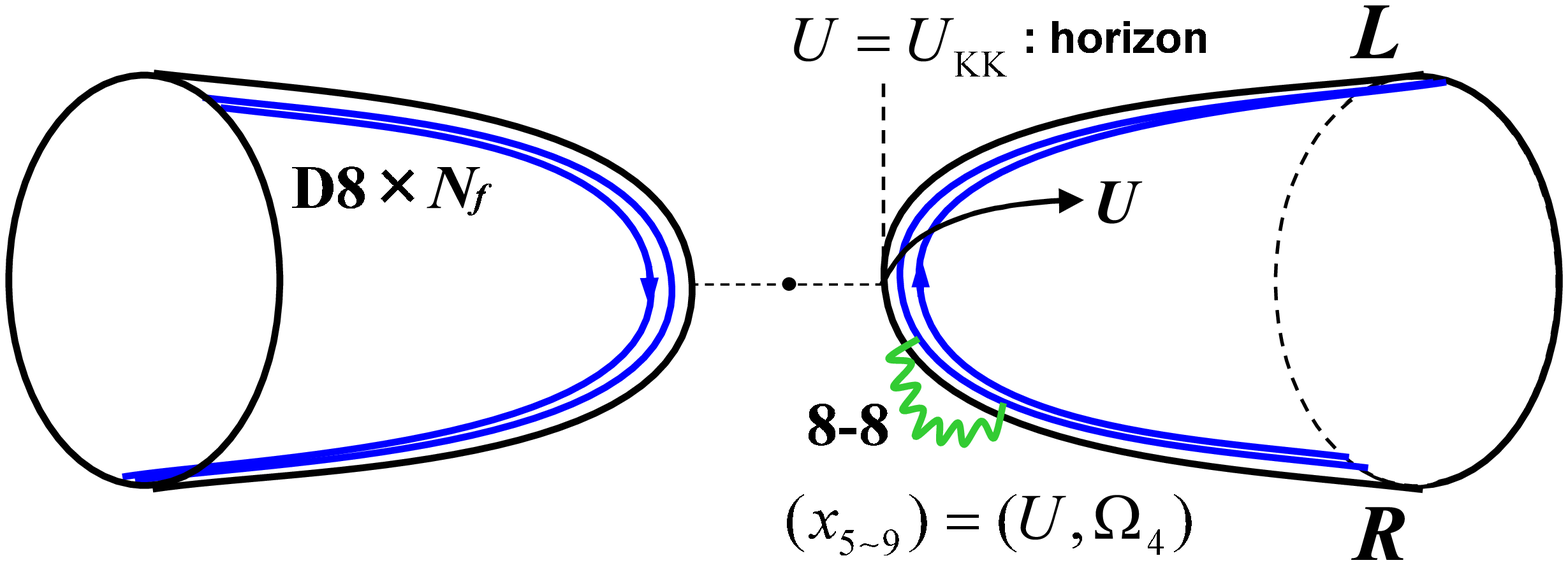}
\vspace{-0.6cm}
       \caption{(a) The multi-D-brane configuration 
corresponding to massless QCD: 
spatially $S^1$-compactified $N_c$ D4-branes attached with 
$N_f$ D8-$\overline{\rm D8}$ pairs. 
Gluons appear from 4-4 strings, and 
quarks appear from 4-8 and 4-$\overline{8}$ strings.  
(b) D8-branes with D4 supergravity background. 
Mesons appear from 8-8 strings.}
\end{center}
\vspace{-1cm}
\end{figure}

Usually, holographic QCD is based on large-$N_c$ argument, 
where $N_c$ D4-branes have a large mass 
proportional to $N_c$. 
With the gauge/gravity duality, 
$N_c$ D4-branes can be replaced by the gravity field, 
and the system becomes $N_f$ D8-branes in the presence 
of the gravity background of the D4-brane, as shown in Fig.1(b).
In large $N_c$ ($\gg N_f$), 
the gravitational contribution from D8/$\overline{\rm D8}$ can be neglected, 
which corresponds to the quenched approximation \cite{NSK06}.

In large $N_c$ and large 't~Hooft's coupling $\lambda$, 
$N_f$ D8-branes in the D4 gravity background lead to 
the Dirac-Born-Infeld (DBI) action in nine-dimensional space-time.
Here, only flavor degrees of freedom remains, since 
the D4-brane with color is already replaced by 
the gravity background.

The DBI action reduces to 
a five-dimensional flavored Yang-Mills theory 
with a curved metric on extra fifth-coordinate, in the leading order. 
The 5-dimensional flavored gauge field $A^\mu(x_\nu,z)$ appears 
from 8-8 strings, 
and the fifth-coordinate $z$ appears from the distance from the D4-brane. 

Through the mode expansion of $A^\mu(x_\nu,z)$ in the fifth $z$-direction, 
a four-dimensional effective theory 
of pions and (axial) vector mesons can be obtained 
from the holographic construction of QCD. 
This meson effective theory has only two independent parameters, and 
describes many phenomenological features of mesons \cite{SS05}.

\section{Baryons as brane-induced Skyrmions}

In large $N_c$, QCD is reduced to weakly interacting theory of mesons 
and glueballs \cite{tH74}, 
and baryons do not appear as direct degrees of freedom.
Instead, the baryon appears as a topological chiral soliton \cite{S62,ANW83}
of mesons in large-$N_c$ QCD. 

In Ref.\cite{NSK06}, we performed the first study of 
the baryon as the chiral soliton in holographic QCD \cite{NSK06}.
As a remarkable fact, 
the topological chiral soliton picture for baryons can be 
directly derived from QCD in the holographic framework: 
the Skyrme term appears from holographic QCD 
without appearance of other four or more derivative terms 
in the leading order of 1/$N_c$ and $1/\lambda$.

For the argument of QCD, the Kaluza-Klein mass $M_{\rm KK}$ plays a role of 
ultra-violet cutoff scale.
Considering the consistency with 
the cutoff scale of $M_{\rm KK}\sim 1{\rm GeV}$ in the holographic approach, 
we take only pions and $\rho$-mesons, and 
derive their four-dimensional effective action \cite{NSK06}
without use of small-amplitude expansion, to describe 
the nonlinear configuration of solitons.

In the chiral soliton picture, 
the baryon is described with the hedgehog configuration 
of Nambu-Goldstone bosons. 
For the chiral field $U(x_\nu) \equiv e^{i\pi(x_\nu)/f_\pi} \equiv 
\{\sigma(x_\nu)+i\tau^a\Pi^a(x_\nu)\}/f_\pi$ 
and the $\rho$-meson field $\rho^\mu (x_\nu)$, the hedgehog configuration 
can be expressed as 
\begin{eqnarray}
& & U^{\star}({\bf x})=e^{i\tau^a \hat{x}^a F(r)}, \nonumber \\ 
& & \rho^{\star}_{i}({\bf x})
= \varepsilon_{i ab}\tau^a\hat{x}^b\tilde{G}(r), \quad 
\rho^{\star}_{0}({\bf x})=0, 
\end{eqnarray}
with $x^\mu=(t, {\bf x})$, $r \equiv |{\bf x}|$ and $\hat x^a \equiv x^a/r$. 
For the baryon with unit topological charge, 
the pion profile $F(r)$ has the topological 
boundary conditions, $F(0)=\pi$ and $F(\infty)=0$. 

We derive the Euler-Lagrange equations, which are coupled nonlinear 
differential equations of pion and $\rho$-meson profiles, 
$F(r)$ and $\tilde G(r)$ \cite{NSK06}. 
Under the topological boundary conditions, we solve the field 
equations and 
obtain the stable chiral soliton solution, 
which we call ``brane-induced Skyrmion'' \cite{NSK06}.
With the experimental inputs of the pion decay constant $f_\pi$=92.4MeV 
and the $\rho$-meson mass $m_\rho$=776MeV 
(i.e., $M_{\rm KK}\simeq$ 948MeV, $e \simeq 7.315$ for the Skyrme parameter), 
we estimate the mass and the radius of the hedgehog baryon as 
$M_B \simeq 834{\rm MeV}$ and 
$\sqrt{\langle r^2 \rangle}\simeq 0.37{\rm fm}$.

\section{Baryonic matter in holographic QCD}

It is interesting to apply holographic QCD to 
finite-density QCD \cite{KSZ07,NSK08}, 
which is difficult to study with lattice QCD.
Here, we consider the baryonic matter in large $N_c$, 
since the holographic QCD is formulated in large $N_c$.

In the large-$N_c$ scheme of the baryonic matter,
the kinetic energy, the N-${\rm \Delta}$ mass splitting 
and quantum fluctuations 
(apart from $O(1)$ zero-point quantum fluctuations) 
are $O(1/N_c)$, 
so that they are suppressed relative to the static baryon mass of $O(N_c)$, 
and the baryonic matter reduces to static solid-like soliton matter, i.e.,
static brane-induced Skyrme matter 
in holographic QCD \cite{NSK08}.

To analyze such static Skyrme matter, we take a mathematical trick 
proposed by Manton and Ruback \cite{MR86}.
To simulate the many Skyrmion system, 
one unit cell shared by a Skyrmion in physical coordinate space 
${\bf R}^3$ 
is compactified into a three-dimensional closed manifold $S^3$
with finite radius $R$, as shown in Fig.2.
The single Skyrmion placed on the surface volume $2\pi^2 R^3$ 
of the manifold $S^3$ corresponds to the finite-density baryonic matter 
with $\rho_B=1/(2\pi^2 R^3)$.

Then, we investigate the single brane-induced Skyrmion on $S^3$ 
with the radius $R$, 
which simulates the baryonic matter 
in large-$N_c$ holographic QCD \cite{NSK08}.
We derive and solve the field equations 
of $F(r)$ and $\tilde G(r)$ 
for the hedgehog soliton on $S^3$, 
with $0 \le r \le \pi R$. 
Here, the topological boundary conditions are 
$F(0)=\pi$ and $F(\pi R)=0$. 

\begin{figure}[hb]
\begin{center}
\resizebox{75mm}{!}{\includegraphics{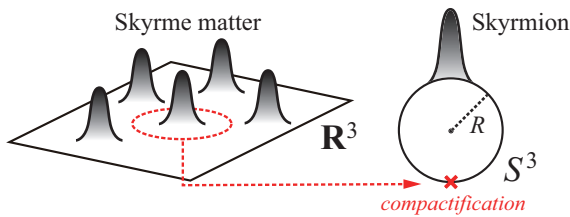}}
\vspace{-0.75cm}
\caption{Schematic figure of the Skyrme matter on a flat physical space 
${\bf R}^3$, and the single Skyrmion on $S^3$ with a finite radius $R$.
The single Skyrmion on $S^3$ simulates 
the Skyrme matter on ${\bf R}^3$.
}
\end{center}
\vspace{-0.2cm}
\end{figure}

\begin{figure}[ht]
\begin{center}
\resizebox{58mm}{!}{\includegraphics{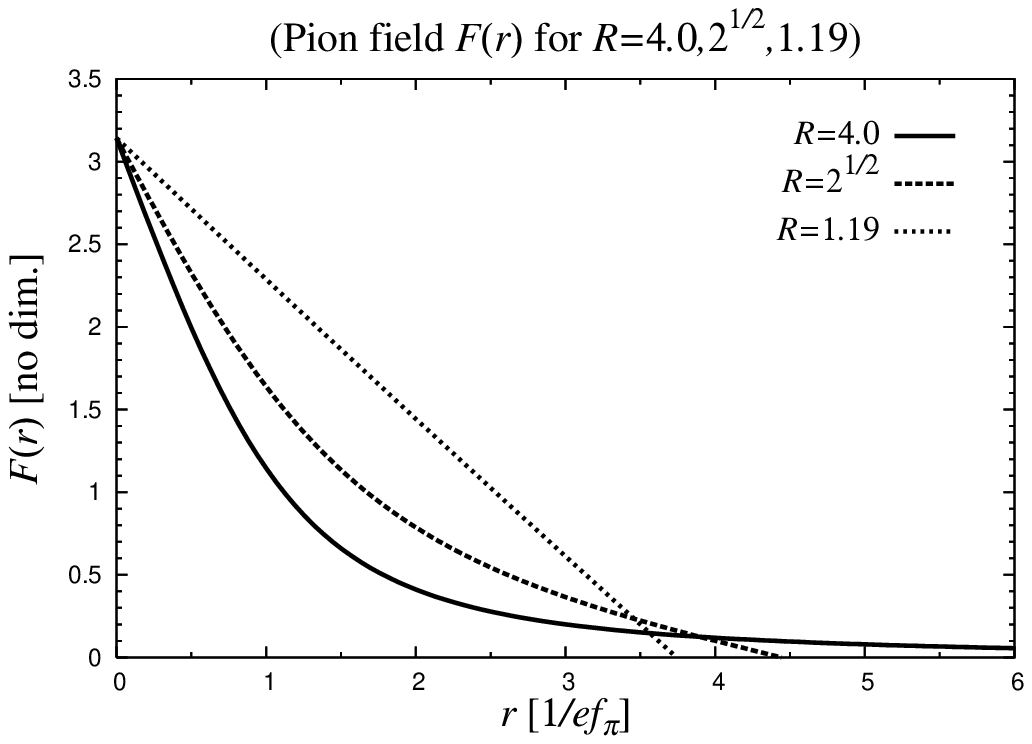}}\\
\vspace{0.48cm}
\resizebox{58mm}{!}{\includegraphics{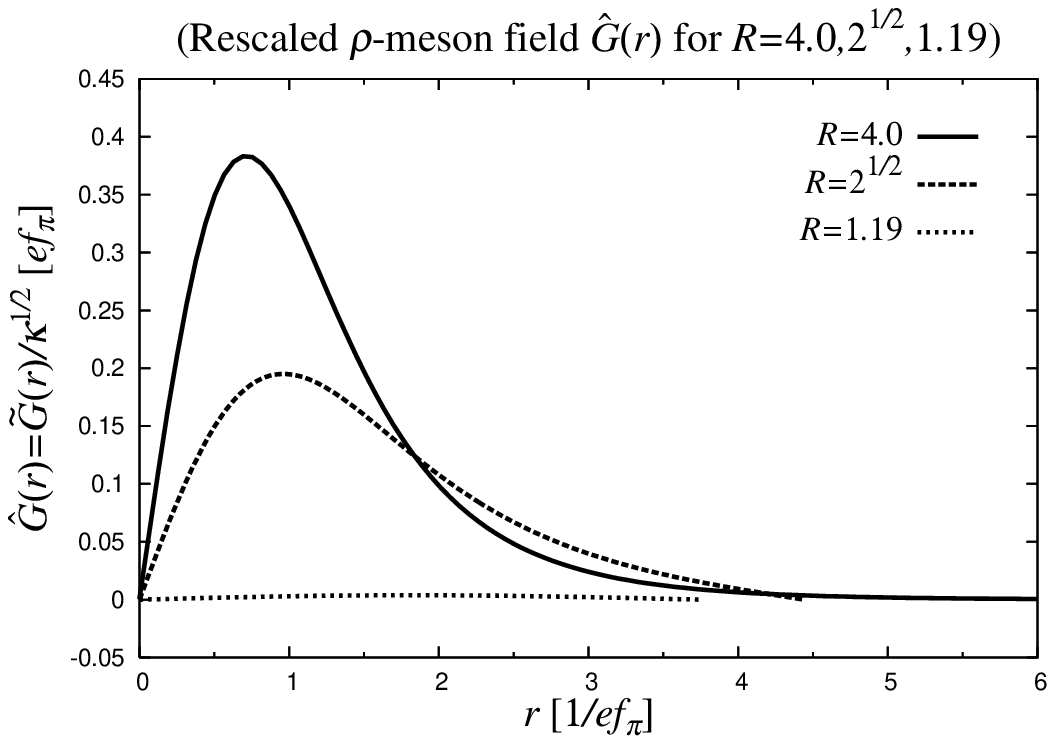}}
\vspace{-0.78cm}
\caption{
(a) The pion profile $F(r)$ and
(b) the $\rho$-meson profile $\tilde G(r)$ 
in the brane-induced Skyrmion on $S^3$ 
with $R$=4.0, $2^{1/2}$, 1.19 [1/$ef_\pi$].
}
\end{center}
\vspace{-0.81cm}
\end{figure}

Figure 3 shows the pion profile $F(r)$ and 
the $\rho$-meson profile $\tilde G(r)$ 
in the brane-induced Skyrmion on $S^3$ with various radius $R$.
As $R$ decreases, 
the $\rho$-meson profile $\tilde G(r)$ decreases, 
and eventually disappears at 
the critical radius $R_c \simeq$ 1.19 [1/$ef_\pi$] corresponding to 
the critical density $\rho_c \simeq 7.12 \rho_0$ \cite{NSK08}.
For $R \le R_c$, only the pion profile $F(r)$ remains, and 
the solution coincides with the ``identity map'', 
$F(r)=\pi-r/R$ 
\cite{MR86}.
Note that the pion field cannot disappear for baryons, 
since the pion profile $F(r)$ has topological boundary conditions, 
unlike the $\rho$-meson profile $\tilde G(r)$.

Figure 4 shows the energy density $\varepsilon(r)$ 
of the single brane-induced Skyrmion on $S^3$ with various values of $R$.
As $R$ decreases, the baryonic soliton is gradually delocalized. 
This delocalization of the chiral soliton on $S^3$ 
indicates the ``swelling of baryons'' in dense matter \cite{NSK08}, 
which leads to the reduction of N-$\Delta$ mass splitting in the framework 
of the chiral soliton. 
For $R \le R_c$, the system becomes homogeneous, and there occurs the 
``delocalization phase transition'' for the baryonic soliton 
\cite{NSK08}, which would suggest deconfinement.

\begin{figure}[ht]
\begin{center}
\resizebox{65mm}{!}{\includegraphics{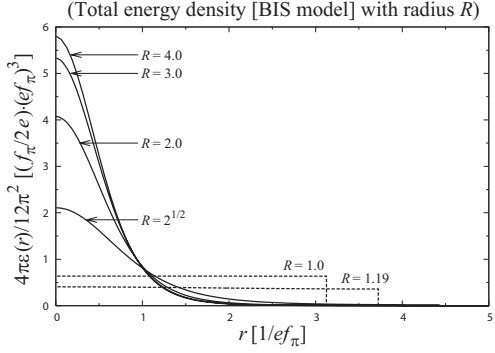}}
\vspace{-0.75cm}
\caption{The energy density $\varepsilon(r)$ 
of the brane-induced Skyrmion on $S^3$ 
with various values of the radius $R$ [1/$ef_\pi$].
The system becomes homogeneous for $R \le R_c (\simeq 1.19)$.
}
\end{center}
\vspace{-1cm}
\end{figure}

For the quantitative description of delocalization 
of the baryonic soliton, we define 
the ``localization order parameter'' $\Phi(R)$ 
from the spatial fluctuation of 
the energy density $\varepsilon({\bf x})$ as 
\begin{eqnarray}
\Phi(R)\equiv\frac{1}{2}\int_{S^3}d^3 x |\bar{\varepsilon}({\bf x})
-\bar{\varepsilon}_{\rm id}|, \quad 
\bar{\varepsilon}({\bf x})\equiv \frac{\varepsilon({\bf x})}{E} 
\end{eqnarray}
with 
$E \equiv \int_{S^3}d^3 x~\varepsilon({\bf x})$ 
and 
$\bar{\varepsilon}_{\rm id}\equiv 1/(2\pi^2 R^3)$.
We also investigate the manifestation of the chiral symmetry 
in the nonlinear representation, 
using the spatially-averaged chiral condensate on $S^3$, 
$\langle \sigma(x_\mu) \rangle_{S^3}$.

As $R$ decreases, both order parameters $\Phi(R)$ and 
$\langle \sigma(x_\mu) \rangle_{S^3}$ decrease, as shown in Fig.5.
At the critical radius $R_c \simeq 1.19$ [1/$ef_\pi$], 
they go to zero, that is, 
the system becomes homogeneous and globally chiral restored, 
which would indicate deconfinement and 
chiral symmetry restoration at high density of $\rho_c\simeq 7.12 \rho_0$.

In summary, we have studied baryons and baryonic matter 
in holographic QCD using a multi-D-brane system in the superstring theory. 
We have obtained the chiral soliton solution of the hedgehog baryon 
for the first time in holographic QCD. 
For the analysis of baryonic matter, 
we have investigated the chiral soliton on $S^3$ in holographic QCD, 
and found the delocalization or the swelling of the baryonic soliton. 
We have found the delocalization phase transition 
at the critical density of about 7$\rho_0$, 
which would indicate deconfinement and chiral symmetry restoration.

\begin{figure}[h]
\begin{center}
\resizebox{60mm}{!}{\includegraphics{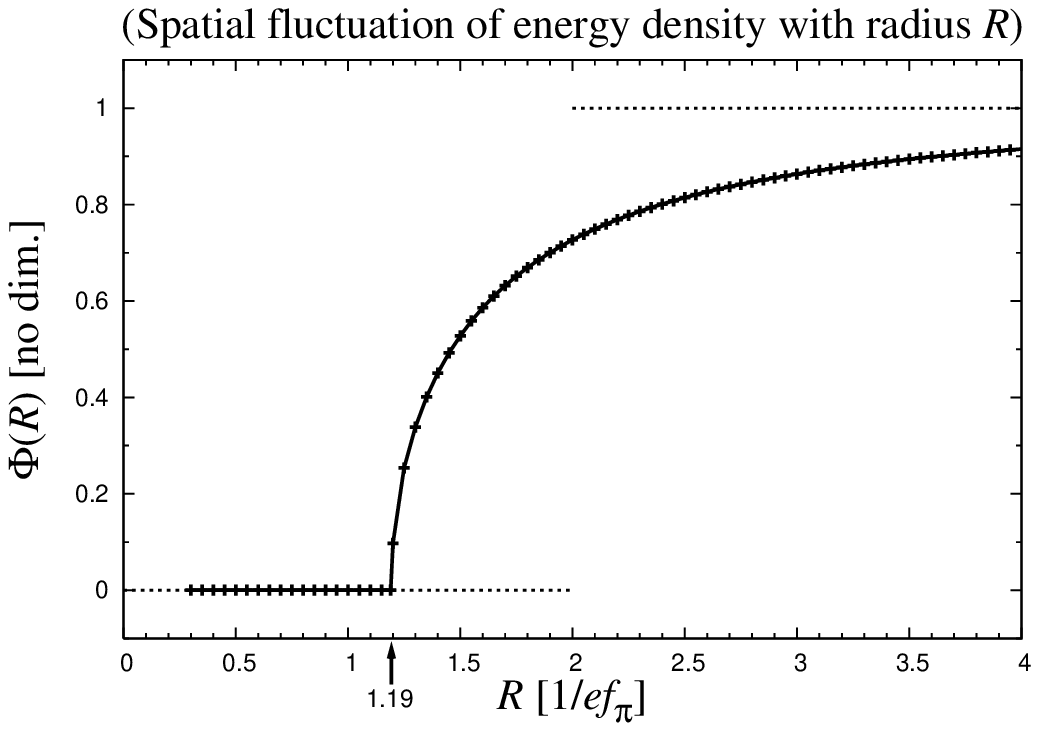}}\\
\vspace{0.4cm}
\resizebox{60mm}{!}{\includegraphics{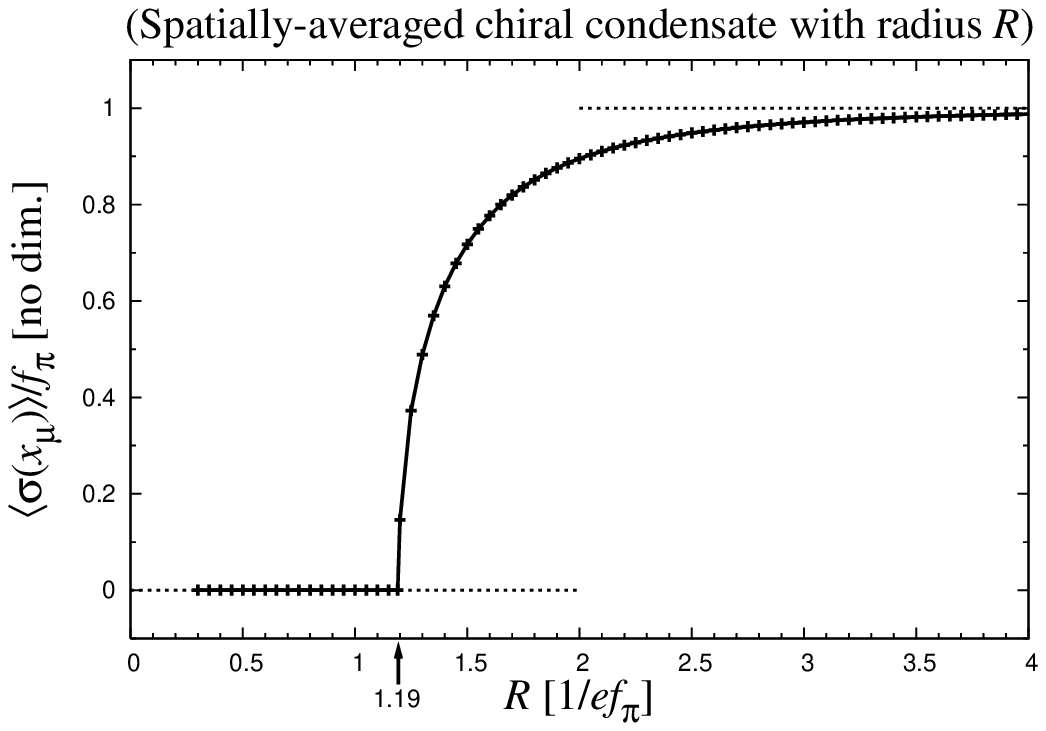}}
\vspace{-0.8cm}
\caption{(a) The localization order parameter $\Phi$ and
(b) the spatially-averaged chiral condensate 
$\langle \sigma(x_\mu) \rangle_{S^3}$ 
as the function of the radius $R$ of $S^3$.
}
\end{center}
\vspace{-1.1cm}
\end{figure}

\end{document}